# Nanostructured superlattices as a probe of fermiology in Weyl-semimetal NbP

Nathan C. Drucker[1*†], Federico Balduini[1], Jules Schadt[1,2], Lorenzo Rocchino[1], Tathagata Paul[1], Vicky Hasse[3], Claudia Felser[3], Heinz Schmid[1], Cezar B. Zota[1†], and Bernd Gotsmann[1†]

1. IBM Research, IBM Research Europe—Zürich, Saümerstrasse 4, 8803, Rüschlikon, Switzerland
2. École Polytechnique Fédérale de Lausanne, Lausanne, Switzerland
3. Max Planck Institute for Chemical Physics of Solids, 01187 Dresden, Germany
† nathan.drucker@ibm.com, zot@zurich.ibm.com, bgo@zurich.ibm.com

**When a charged particle is subject to both a periodic potential and magnetic field, oscillations in its conductivity occur when the field-induced cyclotron radius is commensurate with the period of the potential. This effect has been observed in two-dimensional systems including electron gases and graphene, and is related to the Hofstadter spectrum of magnetic minibands. Here we show that commensuration oscillations also can arise in the ballistic transport regime of a nanostructured 3D material, with the Weyl semimetal NbP. These oscillations encode information about the quasiparticles at the Fermi-surface, including their momentum, charge, mass, and rotational symmetry. In particular, we use the magnetic field and temperature dependence of the commensuration oscillations to extract the Fermi-momenta and quasiparticle mass, in good agreement with Shubnikov-de Haas quantum oscillations. Furthermore, we investigate the relationship between the engineered superlattice symmetry and resistivity response based on the symmetry of the Fermi surface and charge of the quasiparticles. These results demonstrate how nanopatterned superlattices can be used to characterize a 3D materials' fermiology, and also point towards new ways of engineering quantum transport in these systems.**

**Main**

Artificial superlattices exhibit properties that differ from those of their underlying constituents'. This principle has helped guide the architecture of materials and devices across virtually all length-scales, from mechanical metamaterials with macroscopic features[1,2] to photonic crystals[3–7] on the order of micrometers and Moiré superlattices on the nanometer scale[8–13]. An important consideration in the design of superlattices is creating a periodic structure with a periodicity that matches a relevant lengthscale in the underlying medium. For example, in a photonic metamaterial the artificial unit cell period is typically close to the wavelengths of light it is designed to be used in. Similarly, the superlattice period in Moiré materials is small enough to be close to the de-Broglie wavelength of un-twisted materials, resulting in strongly renormalized electronic properties.

One consequence of introducing a periodic potential to an electronic system subject to a magnetic field is that the field induced cyclotron orbit radius can become commensurate with the superlattice potential, leading to oscillations in the material's resistance[14–16]. The additional translational symmetry imposed by the superlattice splits the degeneracy of the Landau Levels, yielding a complicated fractal energy spectrum in magnetic field called Hofstadter's spectrum[8,9,17–19]. Encoded in Hofstadter's spectrum is information about the quantum Hall effect[20–22], as well as other quantum transport phenomena such as Brown-Zak oscillations and geometric commensuration oscillations (COs)[23–25]. Quantum mechanically, COs correspond to oscillations in the superlattice-imposed magnetic miniband width as the flux is changed, leading to oscillations in the magnetic miniband conductivity which are measurable in quantum transport experiments[19,26].

Semi-classically, the commensuration oscillations correspond to cyclotron orbits that get pinned and scattered chaotically by the artificial superlattice[15,16,27,28]. The COs occur when the cyclotron orbits enclose a quantized number $n$ of superlattice unit cells, with $na = 2r_c$. In this expression, $n$ is an integer, $a$ is the lattice spacing of the superlattice, $r_c = \frac{\hbar k_f}{eB}$ is the cyclotron radius, $k_f$ is the Fermi-momentum, $B$ is the magnitude of the applied magnetic field, $e$ is the charge of an electron, and $\hbar$ is the reduced Planck's constant. Commensuration oscillations have been observed in high mobility two-dimensional electron systems with artificially defined superlattices on the order of tens to hundreds of nanometers, including two-dimensional electron gases in III-V comound semiconductor heterostructures[14–16], monolayers of graphene[24,25,29,30], surface states of a topological insulator[31], and incommensurate charge-density waves[32]. In addition to shedding light on the fundamental properties of the Hofstadter spectrum, commensuration oscillations have also been used to prove the fermionic nature of the fractional Quantum hall state[33]. The investigation of commensuration oscillations has so far been limited to 2D materials with isotropic Fermi-surfaces.

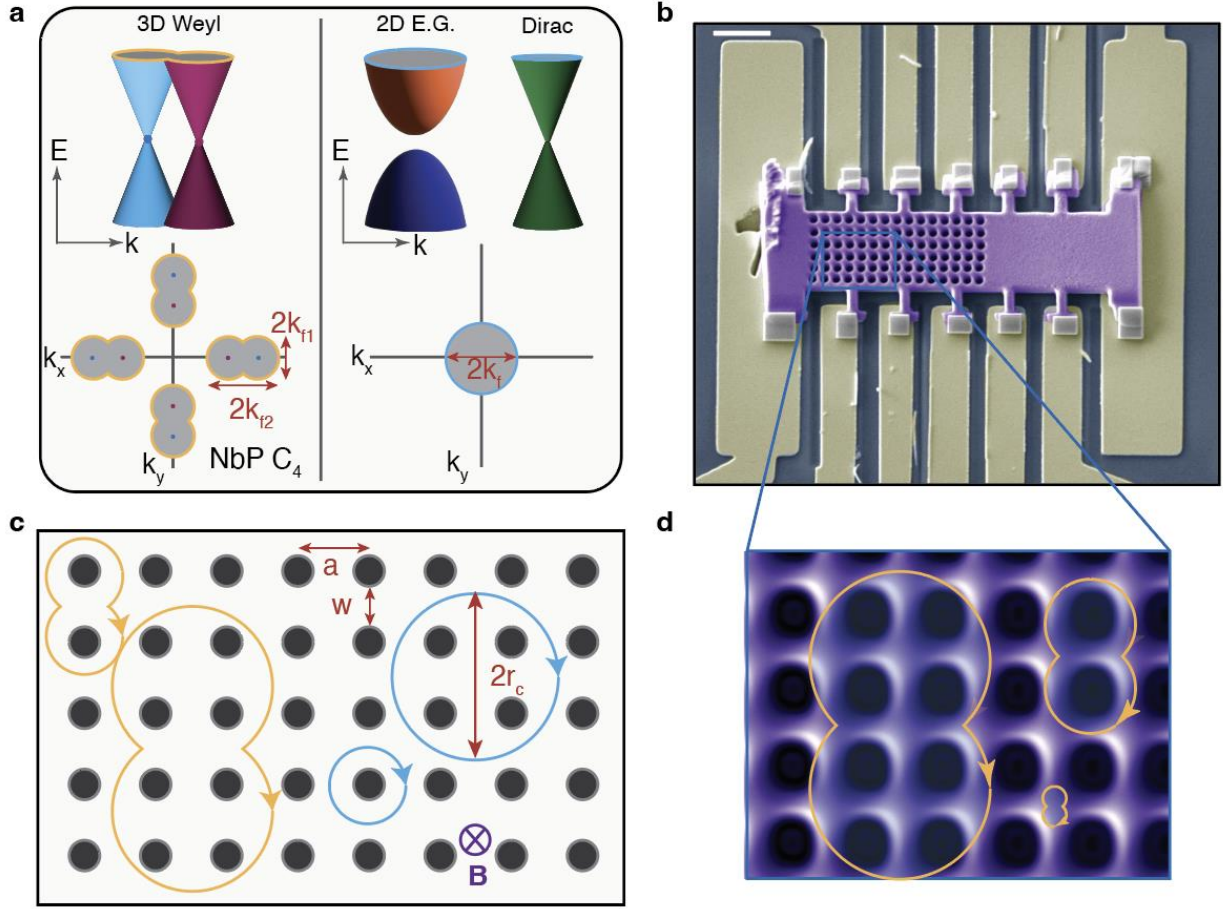

*Figure 1*: **Device design and commensuration principle**. **a** Comparison between the electronic structures in a 3D Weyl semimetal like NbP (left) and 2D electron systems like graphene (right). The Fermi pockets of the Weyl semimetal have a characteristic 'peanut shape' arising from the separation of the Weyl nodes and have two distinct Fermi momenta $k_{f1}$ and $k_{f2}$, while the fermi pockets in a 2D electron gases are isotropic with only one characteristic Fermi momentum $k_f$. In addition, the Fermi pockets in NbP are related to one another by the C4 symmetry of the lattice. Meanwhile, the Fermi pocket of 2DEGs are isotropic. **b** False colored SEM image of a Hall bar made by FIB lithography. Devices include a region with a superlattice on the order of hundreds of nanometers, and an unpatterned region used as a reference. Scale bar is 5 $\mu m$. **c** Depiction of the commensuration effect. With an applied magnetic field out of plane (purple arrow), charged particles precess in the plane of the superlattice (black dots). For a material with an isotropic Fermi surface, such as graphene or a 2D electron gas, the cyclotron orbit is circular (blue). NbP has Fermi pockets with a 'peanut' shape and its cyclotron orbits are shown in gold. The commensuration condition is satisfied when an integer number of superlattice unit cells are enclosed by the cyclotron orbits. **d** Magnification of a FIB defined superlattice with period a=730 nm and three possible NbP cyclotron orbits that satisfy the commensuration condition overlayed.

Here, we use focused ion beam (FIB) machining to engineer nanoscale antidot superlattices in 3D topological Weyl semimetal NbP. We observe commensuration oscillations with a characteristic field dependence that persist to high temperature (T~150K). To demonstrate that similar information about the Fermi surface can be attained from quantum oscillations (QOs) and COs, we use NbP because of its long mean free path. NbP has a mean free path $l_{mfp} \sim 10 \ \mu m$ at liquid Helium temperatures[34] which is longer than the lithographically defined superlattice period ($l_{mfp} > a$). By changing the rotational symmetry of the superlattice we quench the strength of the CO peaks, demonstrating that their intensity reflects the in-plane rotational symmetry of the Fermi surface. We further introduce a chiral superlattice that has an asymmetric response to the magnetic field, dictated by the negative charge of the ballistic quasiparticles. Finally, we infer the out-of-plane structure of the Fermi-surface by tracking the dependence of the CO peaks on the angle of the applied magnetic field. The effective mass and out-of-plane Fermi surface shape are complemented by temperature- and angular- dependent Shubnikov-de Haas (SdH) measurements, showing methodological consistency but also clearly demonstrating the wealth of information contained in the CO peaks compared to conventional QOs. Apart from demonstrating antidot superlattices as a powerful tool to characterize the electronic properties of quantum materials, we illustrate how specific electronic responses may be engineered by tuning the relative lengths scales and symmetries of both the underlying 3D quantum material and superlattice potential.

**Magnetotransport in Weyl-semimetal superlattices**

Weyl semimetals are ideal materials in which to observe commensuration oscillations because of their large mean free paths and relatively small Fermi surfaces[35]. These properties make COs observable under the constraints of superlattice periods achievable with FIB machining[36] and lab-based magnetic fields of a few Tesla. Unlike graphene and 2DEG systems, which have isotropic circular Fermi surfaces, NbP has a four-fold symmetric Fermi surface, with each pocket constituted by two slightly overlapping lobes[34,37] (Figure 1a). The non-circular nature of these Fermi-pockets comes from the separation of the Weyl nodes. As the real-space cyclotron motion is related to the reciprocal-space shape of the Fermi surface, the non-circular Fermi pockets change the commensuration condition. For example, in a square antidot lattice -- like the one shown in

Figure 1c -- the anisotropic Fermi surface of NbP heuristically leads to a modified commensuration condition

$$na = 2r_{c1} = \frac{2\hbar k_{f1}}{eB}, \qquad 2ma = 2r_{c2} = \frac{2\hbar k_{f2}}{eB} \qquad (1a, b)$$

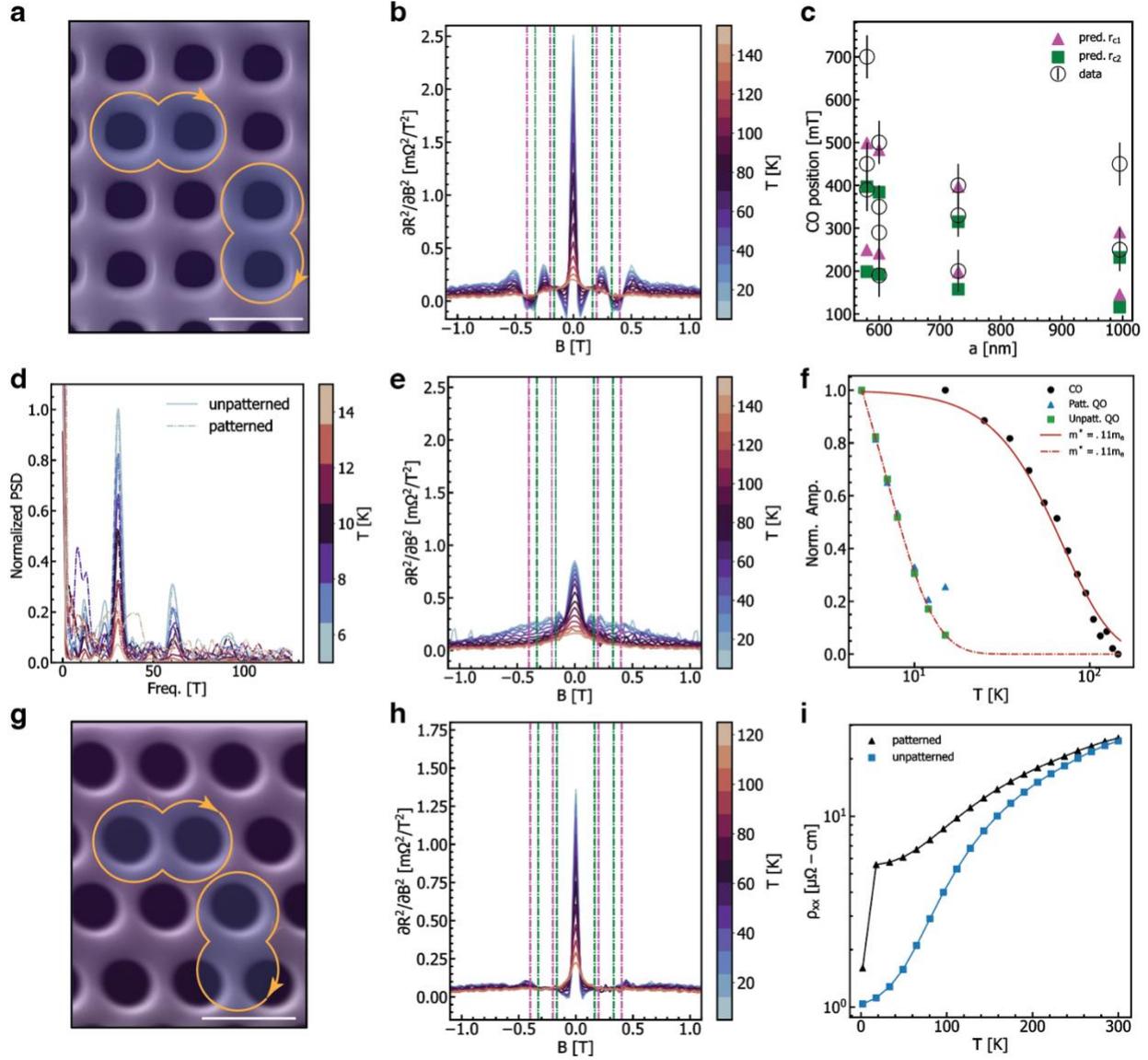

*Figure 2* **Field and temperature dependence of commensuration oscillations in longitudinal resistance. a** False colored SEM image of NbP patterned with a square superlattice, with a = 730 nm. Yellow paths are representations of the cyclotron orbits in real space. Scale bar in bottom right is 1 $\mu m$. **b** Second derivative of resistance with respect to magnetic field for the square superlattice device. Minima in the second derivatives correspond to peaks in resistivity, helping to identify commensuration peaks at B =.33 T and B = .40 T (green and magenta dashed lines, respectively). **c** Resistivity peak position in magnetic field for each of the four 4-fold symmetric superlattice devices in this study with different superlattice spacings. Green squares correspond to predictions of Eq. 1a whereas magenta triangles correspond to the predictions of Eq. 1b. Black open circles correspond to measured

commensuration oscillations. **d** Normalized Fourier amplitude of the high field quantum oscillations for the patterned and unpatterned devices, (dashed and solid lines, respectively) for temperatures between 5 K and 15 K. The spectra of both sides have primary peaks around 31 T and 61 T, consistent with literature. **e** Second derivative of resistance with respect to magnetic field for the square unpatterned device, showing no CO peaks. **f** Temperature dependence of the COs (black points) and QOs (blue for patterned, green for unpatterned). Black points are extracted from the second derivative minima in panel **b**. Red line is a fit to the modified Lifshitz-Kosevich formula and extracted $m_{co}^* = .11 m_e$. **g** False colored SEM image of NbP patterned with a triangular superlattice, with a =730 nm. Yellow paths are representations of the cyclotron orbits in real space. Scale bar in bottom right is 1 um. **h** Second derivative of resistance with respect to magnetic field for the triangular superlattice device, which has quenched CO features. **i** Temperature dependence of the longitudinal resistivities of the patterned (black triangles) and unpatterned (blue squares) sides. Below T ~ 4 K, the amorphous surface layer of Nb becomes superconducting, leading to the resistivity drop in the black curve.

For $k_{f1} = .022 \frac{1}{Å}, k_{f2} = .035 \frac{1}{Å}$, [34] and $m, n$ are integers. An example of a typical device is shown in Figure 1b, which has a superlattice pattern defined on one side, such that the transport can be readily compared to the device's unpatterened side. The commensuration oscillation conditions are depicted in Figure 1c, and superimposed on a magnification of the superlattice in one of our devices in Figure 1d. These conditions show how introducing a more complicated Fermi surface affects the commensuration oscillations. The symmetry of the Fermi-surface imposes conditions on commensuration oscillations that strongly contrast with previously investigated isotropic systems.

We observe commensuration oscillations in the longitudinal resistivity of the our superlattice devices for magnetic fields perpendicular to the current direction. Figure 2a shows an SEM image of a square superlattice device with spacing a = 730 nm and Figure 2b shows the second derivative of this device's resistance with respect to magnetic field. Valleys in the second derivative correspond to peaks in the resistivity, helping to identify the commensuration oscillations. From the second derivative of the resistance with respect to field, we identify commensuration peaks at $B_1 = .33$ T and $B_2 = .40$ T, which correspond to the first harmonic commensuration conditions given by Equation 1a,b. These magnetic field values are in excellent agreement to the commensurability condition expected for a lattice period $a = 730$ nm and the Fermi-momenta of NbP known from prior studies[34]. To confirm the validity of the heuristic commensuration conditions in Eq. 1, we fabricate four devices with different superlattice spacings, and find good agreement between the predicted CO peak positions and the lattice spacings (Figure 2c). Conversely, if the Fermi-momenta were unknown, they could be determined by the field dependence of the COs. The values of the Fermi-momenta extracted from the CO peak positions are also consistent with the area of the Fermi surface found with Shubnikov-de Haas oscillations (Figure 2d). Besides serving as a comparison for the Fermi momenta values to the CO peaks, the QO spectrum shows that the superlattice patterning process does not destructively alter the underlying Fermi surface properties of NbP.

Notably, the CO peaks are observable at temperatures an order of magnitude higher than Shubnikov- de Haas QOs, and we use the CO temperature dependence to extract the quasiparticle mass. Conventional quantum oscillations are typically observed at low temperatures because the thermal broadening of the Fermi-Dirac distribution suppresses the magnitude of the quantum oscillations. Commensuration oscillations follow a temperature dependence similar to the Lifshitz-Kosevich form $A = \frac{T/T_c}{\sinh(T/T_c)}$ where $T_c = \frac{\hbar eB}{2\pi^2 k_B m^*}$, but renormalized by a factor of $k_f a/2$ such that $T_c^{CO} = \frac{\hbar eB}{4\pi^2 k_B m^*} k_f a$ [38,39]. This factor reflects how the thermal smearing of the cyclotron orbits impacts the commensuration condition. The amplitude of the commensuration oscillations as a function of temperature — extracted from the minima in Figure 2b — is plotted in Figure 2e. We fit the data with the modified Lifshitz-Kosevich formula (solid red curve) to extract the mass of the quasiparticles from the amplitude change of the commensuration oscillations. The extracted quasiparticle mass m* = .11$m_e$ is in excellent agreement with the mass extracted from the QO (dashed red line in Figure 2e), and literature values for NbP[34,40–42]. Notably, the QOs are visible up to T ~15 K, while the COs are visible up to T ~150 K, demonstrating both the robustness of the CO effect and how they can be used to extract the same information as the QO at more relaxed (higher T, lower B) conditions.

Next, we demonstrate that COs contain additional information about the Fermi surface beyond quasiparticle mass and Fermi momentum. In particular, changing the relative symmetry of the superlattice to that of the underlying crystal provides information about the Fermi-surface symmetry. The Fermi-surface of NbP has four-fold rotational symmetry in the xy-plane due to its tetragonal crystal structure. It is interesting to consider whether different superlattice symmetries impact the commensuration oscillations. To investigate this symmetry-selection, we pattern a triangular (three-fold symmetric) superlattice to see how the CO amplitude changes (Figure 2g).

Indeed, we observe that the CO amplitude is quenched in the three-fold symmetric superlattice (Figure 2h) relative to the four-fold symmetric lattice (Figure 2b). This symmetry selection can be explained by the fact that the triangular superlattice no longer has the same underlying symmetry as the Fermi surface, and thus the commensuration condition is more difficult to satisfy. There are still some signatures of commensuration effects in the triangular superlattice, which may arise from commensuration effects in one direction, but not both. Thus, information about the relative symmetry between the superlattice and Fermi-surface is imprinted on the CO spectrum.

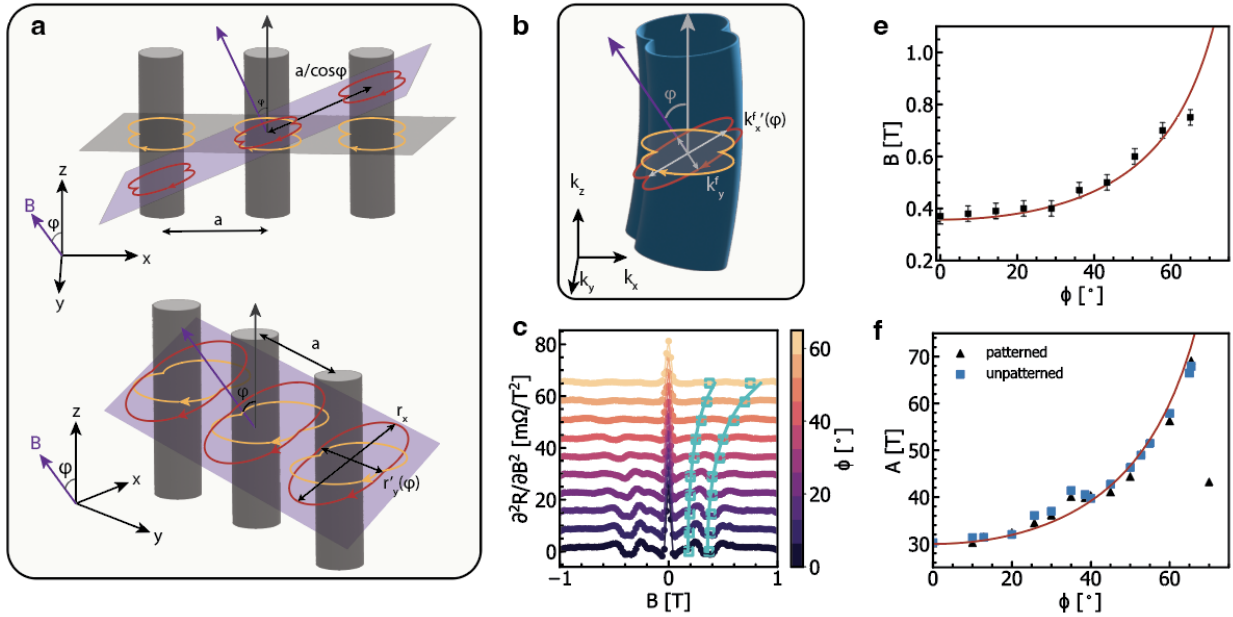

*Figure 3* **Inferring 3D structure of the Fermi surface from commensuration dependence on magnetic field angle. a** Real-space schematic of transport geometry with magnetic field applied at angle $\phi$ in the x-z plane. In the plane of the field rotation (top), the superlattice spacing between the potentials is modified along the x-direction, and the cyclotron radius is unchanged. In the plane perpendicular to the field (bottom), the superlattice spacing is unchanged, but the cyclotron radius in the y-direction becomes modified based on the change of the Fermi-surface. **b** In reciprocal space, the fermi-surface cross section depends on the angle of the field. For a magnetic field in the x-z plane, the fermi-momentum along the $k_x$-direction becomes renormalized and the fermi-momentum along the $k_y$-direction remains unchanged, leading to a change in the cyclotron radius along the y-direction in real space (Figure 4a, bottom). **c** Waterfall plot of second derivative of resistance with respect to field for different field angles in the x-z plane relative to the antidot superlattice. Cyan points correspond to extracted CO positions, and the solid line corresponds to a $1/cos\phi$ fit. **e** Extracted n = 1 CO position as a function of field-angle $\phi$ with solid red line corresponding to fit to the CO data of $1/cos\phi$. **f** Fermi-surface area extracted from the quantum oscillations on both the patterned and unpatterned sides of the device as a function of field angle. Solid red line corresponds to fit to the QO data of $1/cos\phi$.

*Probing the 3D Fermi surface with commensuration oscillations*

In addition to modifying the superlattice symmetry to identify the rotational symmetry of the Fermi-surface, we also gain information about the 3D geometry of the Fermi-surface from the dependence of the CO peaks on the relative angle between the magnetic field and superlattice. One important property of a nanostructured superlattice in a 3D material by comparison to a 2D material is that the 3D material (NbP in this case) has an additional spatial component to its electronic structure. Following the semiclassical equations of motion for a Bloch electron in a magnetic field, the cyclotron orbit of a quasiparticle in a magnetic field is in the plane perpendicular to the applied field[43]. In typical QO measurements, this means that by changing the angle of the applied field, the frequency of the QO changes based on the change in the cross-sectional area of the Fermi-surface perpendicular to the magnetic field. Thus, angle dependent QO measurements are frequently employed to map out the full 3D shape of a material's Fermi surface.

The angular dependence of the cyclotron orbit is also imprinted on the commensuration oscillations, because the radius of the cyclotron orbit in real space dictates the commensuration condition. In real space, the cyclotron orbit changes size in the direction orthogonal to both the magnetic field and the momentum change (Figure 3a). As depicted in Figure 3b, the Fermi momentum changes in the direction orthogonal to the magnetic field, confined to the plane of rotation and acquires angular dependence $k_f(\phi)$. As a result, the commensuration acquires an angular dependence $na = 2r_c(\phi)$. Figure 3c shows the change of commensuration peaks as the field is rotated away from the *z*-direction (out of plane relative to the superlattice). The position of the commensuration peak changes, as does the amplitude. In figure 3e, we plot the dependence of the *n* = 1 commensuration peak as a function of field angle $\phi$. The angular dependence of the CO peak closely follows the $1/\cos\phi$ dependence characteristic of a cylindrical Fermi surface, as fit by the solid line. The cylindrical Fermi surface is a good approximation in NbP based on the anisotropic (but closed) Fermi-surface from DFT calculations for this material[42], although we note that NbP's Fermi surface is closed. The angular dependence of the CO peaks also matches the angular dependence of the Fermi-surface area, as shown in Figure 3f. This comparison demonstrates that the COs can reflect the same angular information as the QOs, but at much lower fields.

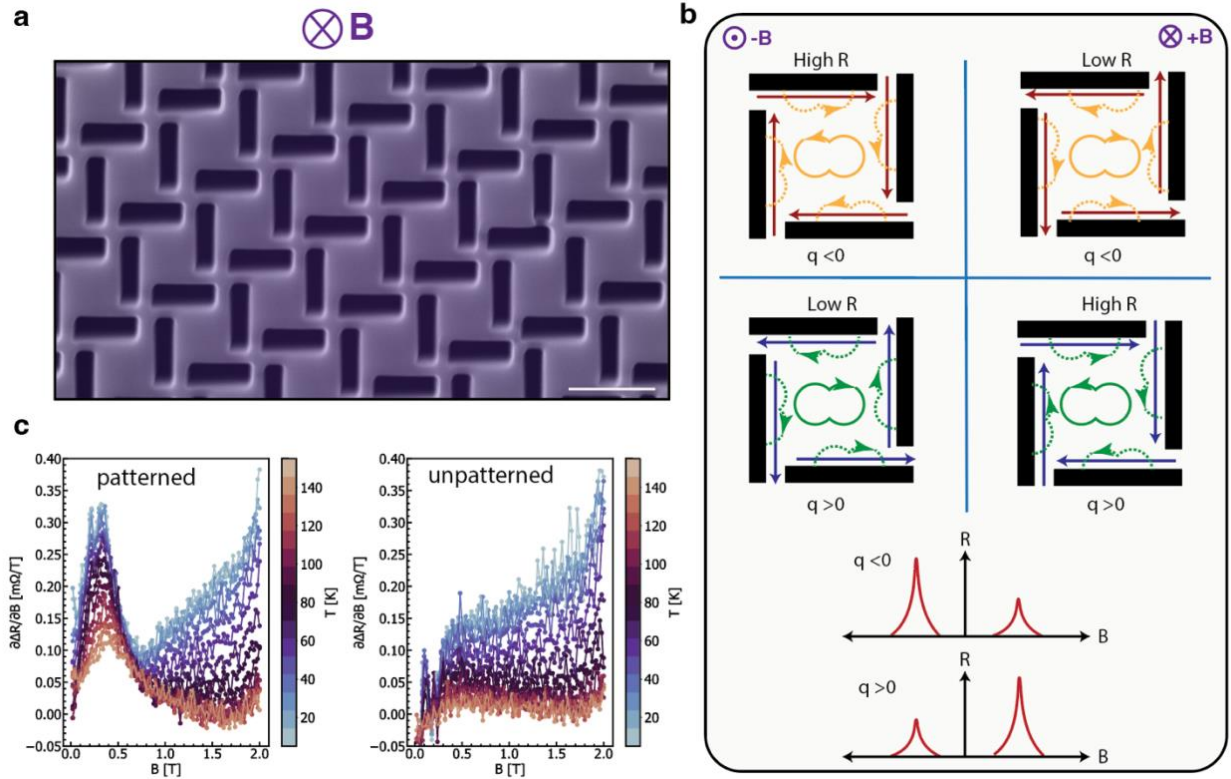

*Figure 4* **Deducing quasiparticle charge from a chiral superlattice**. **a** False colored SEM image of chiral superlattice with fourfold rotational symmetry. Scalebar in bottom right corner is 1 $\mu m$. Positive magnetic field B point into the plane of the superlattice. **b** Cyclotron orbits for oppositely charged particles (yellow for negative, green for positive) are confined to chiral plaquettes at the commensuration condition. At the boundaries of the plaquettes are chiral skipping orbits (dashed lines) whose drift direction is indicated by the red and blue arrows. The relative chirality drift velocities and the boundaries leads to higher or lower resistance peaks, shown in the bottom for positive and negative charges, respectively. **c** First derivative with respect to field of the asymmetric resistance $\Delta R = R(-B) - R(+B)$ for the patterned and unpatterned sides of the chiral device. For the chiral pattern, there is a strong asymmetric response at low field, indicating a negative charge of the quasiparticles based on the chirality of the superlattice.

### *Deducing quasiparticle charge from a chiral superlattice*

Beyond changing the rotational symmetry of the superlattice to characterize the rotational symmetry of the Fermi surface, we extract the quasiparticle charge through the asymmetry of COs in a chiral superlattice. As shown in Figure 4a, the chiral superlattice breaks in-plane mirror symmetry in both the x- and y-directions, resulting in a designated handedness. The chirality of the superlattice becomes imprinted on the magnetoresistance based on the interaction between charge carriers and the superlattice potential. The chirality of the boundary scattering impacts the CO peaks depending on the charge of the quasiparticles. Microscopically, this effect comes from considering that positively and negatively charged quasiparticles have cyclotron orbits with opposite direction for a given magnetic field direction (green and yellow solid orbits in Figure 4b, respectively). At boundaries of the plaquette, cyclotron orbits induce chiral skipping modes. In Figure 4b, the drift direction is represented by the blue and red arrows for positively and negatively charged quasiparticles, respectively. Crucially, in a superlattice with a given chirality, the skipping orbit drift can either flow into a boundary or out of the chiral plaquette. When the direction of the drift is towards the boundary, the resistance is higher from increased boundary scattering. Changing the magnetic field direction or the quasiparticle charge flips the direction of the skipping orbit drift, leading to higher or lower resistance at the CO condition (Figure 4b, bottom).

Hence, by measuring the resistance through the chiral superlattice with both positive and negative magnetic field, we can determine the charge of the quasiparticles by the asymmetric component of the resistance in field $\Delta R = R(-B) - R(+B)$. Figure 4c shows the derivative of the asymmetric resistance $\partial \Delta R / \partial B$ measured for different field strengths, and displays a clear peak in the low field region, where cyclotron orbits become confined to the chiral plaquettes. The amplitude of the peak decreases as temperature is increased, as is characteristic for the ballistic CO effect. We compare the asymmetric response of the chiral device to the unpatterned region of the same sample, which has no equivalent peak (Figure 4c, right). The sign of this peak is consistent with electron-like charge carriers with negative charge interacting with the chiral superlattice in our devices, as depicted by the yellow orbits in Figure 4b. The negative sign of the charge displayed in the chiral superlattice device is also consistent with prior studies of NbP, which show an electron-like nature of the Weyl-fermion pocket[40].

**Fermiology with Commensuration Oscillations vs Quantum Oscillations**

The overlap in information attained from the commensuration oscillations and quantum oscillations is deeply rooted in the similarity of their quantization conditions. The Lifshitz-Onsager quantization rule underpins conventional Shubnikov-de Haas oscillations. Equivalently, Landau levels can be considered Landau 'tubes' in reciprocal space, which cause conductance oscillations when their quantized circumference intersects the cross section of the Fermi surface[43]. These relations yield:

$$(\nu + \lambda) \frac{2\pi eB}{\hbar c} = A_{\perp B}(E_f) \qquad (2)$$

where $\nu$ is the quantum number of the Landau levels, $\lambda$ is a system specific phase correction, B is the magnetic field, and $A_{\perp B}(E_f)$ is the area of the Fermi surface perpendicular to the field. In the antidot case, the quantization condition comes from the real-space condition that an integer number of superlattice unit cells becomes enclosed by the cyclotron radius (Figure 1). In this framing, it is natural that each of these methods provides similar information about the cyclotron orbits.

Engineering the real-space symmetry of the nanostructured superlattice provides an opportunity to attain information about the Fermi-surface beyond what is accessible with conventional quantum oscillations measurements. Quantum oscillations inherit the symmetry of the Fermi pocket whose extremal area satisfies the Onsager quantization condition (Eq. 2), but the Onsager quantization condition does not provide information about the symmetry relations *between* different Fermi-pockets. In contrast, the real-space commensuration condition depends implicitly on the spatial overlap between the cyclotron orbits and the patterned superlattice; if there is not a symmetric match between the symmetry of the superlattice and the crystal structure, the commensuration condition is not satisfied for all cyclotron orbits. We note that prior experiments in 2D systems have not been able to detect this effect because of the isotropic nature of their Fermi-surfaces. Further breaking the mirror symmetry of the superlattice by patterning a chiral motif enables the determination of parity of the quasiparticle charge, which is not information contained in quantum oscillations.

Based on the large lengthscale of the lattice compared to the Fermi-wavelength of NbP, our experiments are limited to the semi-classical, as opposed to quantum, regime. Consequently, the symmetry of the superlattice does not necessarily modify the wavefunction of the quasiparticles at the Fermi surface. In the quantum regime, when the superlattice modulation is on a similar order of magnitude as the Fermi-wavelength, it has been observed that additional quantum oscillations become present on top of the commensuration oscillations[16]. To endow the Weyl electrons with the symmetry of the superlattice and induce new phases of matter, it may be necessary to have the superlattice modulation on a scale of tens of nanometers, matching the Fermi wavelength (for NbP, $\lambda_f \sim 20\ nm$). Achieving such small superlattice sizes remains challenging, as traditional lithographic techniques are not generally applicable to 3D structures like bulk single crystals, and the FIB technique is limited by resolution[36] at such small feature sizes.

**Conclusions**

Extending methods of engineered superlattices to 3D materials requires understanding the fundamental interactions between the superlattice and underlying electronic degrees of freedom. We have explored the properties of the Fermi surface that become imprinted on the magneto-transport of nanostructured NbP, based on the superlattice spacing and symmetry. The commensuration oscillations of resistivity in magnetic field provide a more detailed probe of the Fermi-surface, and at temperatures and magnetic fields more amenable, than quantum oscillations. Beyond demonstrating powerful characterization capabilities, these experiments also point to how unique transport effects can be engineered based on the relative properties of the Fermi-surface and the nanostructured superlattice. For example, it is promising to combine phonon engineering[44–46] with electronic superlattice engineering in Weyl semimetals, especially given their unique thermal and thermoelectric transport properties[47–50]. As topological semimetals find more applications[51], it is attractive to consider nanostructured superlattices as a means of creating properties on demand in this novel materials class.

**Methods**

*NbP Crystal Growth*
Polycrystalline NbP powder was synthesized by reacting Nb powder (99.9% purity) and red phosphorous pieces (99.999% purity) in an evacuated silica tube for 48 h at 800 °C. Single NbP crystals were then grown from this powder through a chemical transport reaction using iodine as the transport agent, with the source and sink set at 950 °C and 850 °C, respectively.

*Device Fabrication*
Microstructured single crystals of NbP were shaped using a FEI Helios 600i Ga+ Focused Ion Beam system, using an ion accelerating voltage of 30 kV. When patterning the superlattices, a current of 80 pA was used. The lamella were extracted from the single crystal samples, shaped to the desired specifications, and then placed on a prefabricated substrate within the vacuum of the FIB. These substrates consist of sputtered gold pads on top of Silicon, with an insulating oxide layer in between. Contact between the gold pads and FIB device was made with the FIB Pt.

*Magnetotransport Measurements*

AC electrical transport measurements were made by applying constant current along the *a*-direction of NbP and measuring voltage along transverse and longitudinal directions with respect to the current. These measurements were done in a PPMS Dynacool cryostat with external lock-in amplifiers (MFLI, Zurich Instruments). Magnetic field was applied along the *c*-direction unless otherwise specified.

**Acknowledgements**

We wish to acknowledge the support of the Cleanroom Operations Team of the Binnig and Rohrer Nanotechnology Center (BRNC). We also thank Jonathan Eroms at University of Regensburg for helpful discussions. This work was supported by the Horizon Europe project AttoSwitch, the Swiss State Secretariat for Education, Research and Innovation (SERI) under contract number (SERI contract no. 23.00594).

**Competing Interests**

The authors declare no competing interests.

# Supplementary Information: Nanostructured superlattices as a probe of fermiology in Weyl-semimetal NbP

Nathan C. Drucker[1*†], Federico Balduini[1], Jules Schadt[1,2], Lorenzo Rocchino[1], Tathagata Paul[1], Vicky Hasse[3], Claudia Felser[3], Heinz Schmid[1], Cezar B. Zota[1†], and Bernd Gotsmann[1†]

1. IBM Research, IBM Research Europe—Zürich, Saümerstrasse 4, 8803, Rüschlikon, Switzerland
2. École Polytechnique Fédérale de Lausanne, Lausanne, Switzerland
3. Max Planck Institute for Chemical Physics of Solids, 01187 Dresden, Germany
† nathan.drucker@ibm.com, zot@zurich.ibm.com, bgo@zurich.ibm.com

## S1: Quantum oscillations

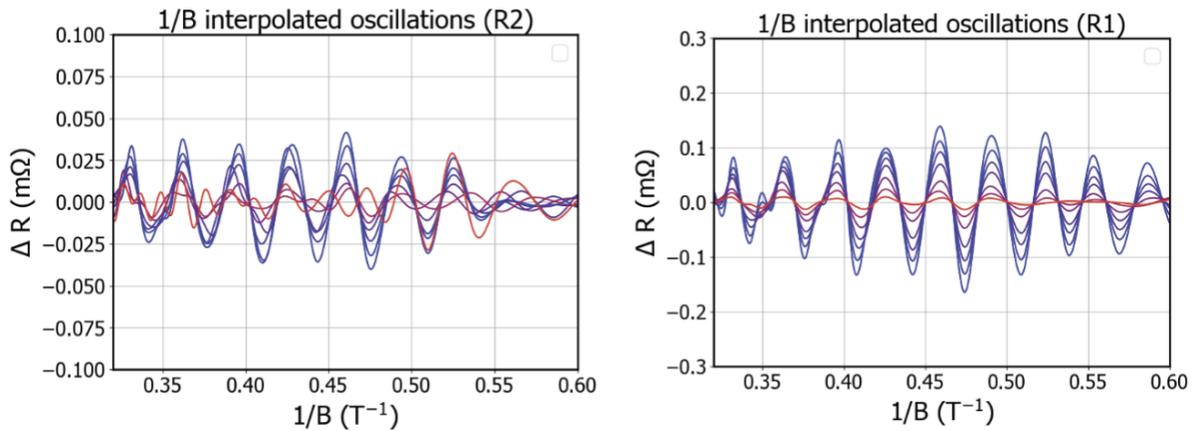

*Figure S1* **Isolating quantum oscillations from raw data**. (left) Quantum oscillations in the antidot device's Resistance as a function of 1/B. (right) Quantum oscillations in the unpatterned device's Resistance as a function of 1/B. Different color plots correspond to temperatures varying from T=2K (Blue) to T=15K (Red).

Quantum oscillations were extracted from the raw magnetoresistance curves by applying a Savitzky-Golay filter with large averaging window and low polynomial order, and then subtracting the result from the raw data. The isolated oscillations were then smoothed with a third order spline, with the smoothed oscillations shown above. To attain the area of the Fermi surface associated with the oscillations, we take the Fourier transform of the oscillations as a function of (1/B) and end up with the power spectral density (PSD) that is plotted in Figure 2 of the main text.

## S2: Commensuration oscillation conditions

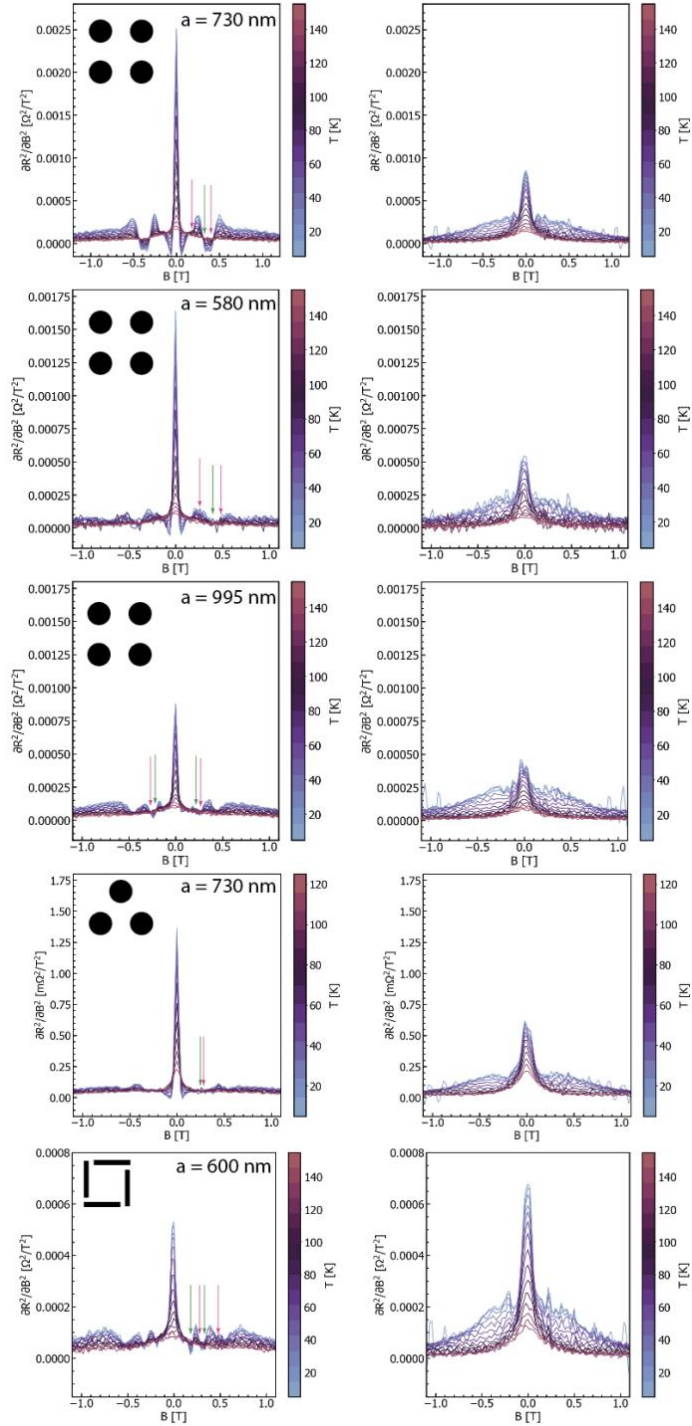

*Figure S2* **Identifying commensuration peaks in different devices**. Comparison of transport in five different devices, with varying lattice spacing and superlattice symmetry. Each row shows second derivative with respect to magnetic field for patterned (left) and unpatterned (right) sides of the devices. Pink arrows indicate the location of commensuration peaks

**S3: Triangular vs Square superlattice Hall data**

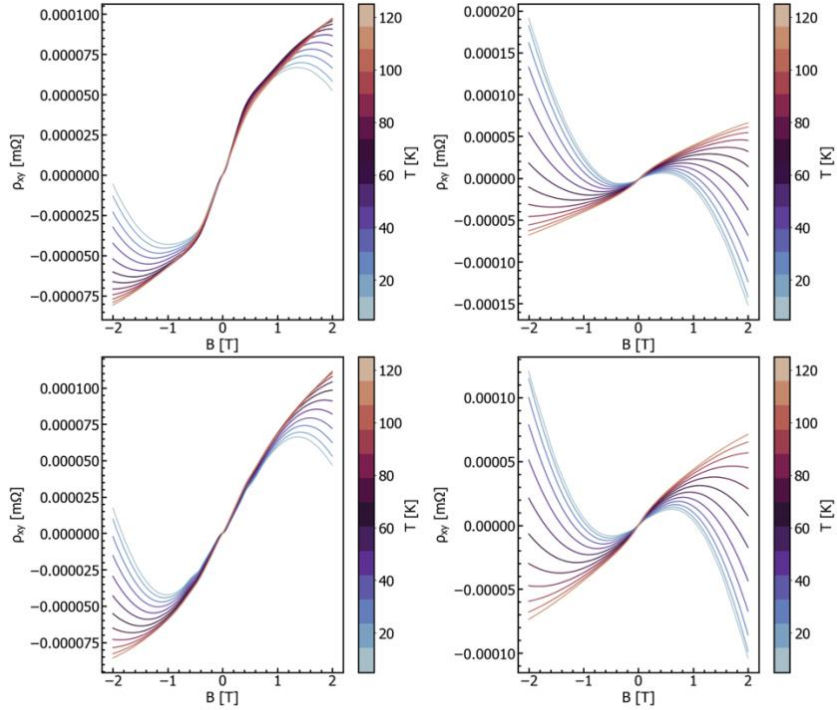

*Figure S3* **Hall response of triangular and square superlattices**. Transverse resistivity for the triangular (top) and square (bottom) superlattice devices. Left column is patterned side of devices, right column the unpatterned side. Based on the similarity between the two devices' Hall response, we conclude that the lack of commensuration effects in the triangular device comes from a difference in symmetry, not from a difference in carrier concentration or mobility, from the square superlattice devices.

## S4: Asymmetry of chiral vs square lattice

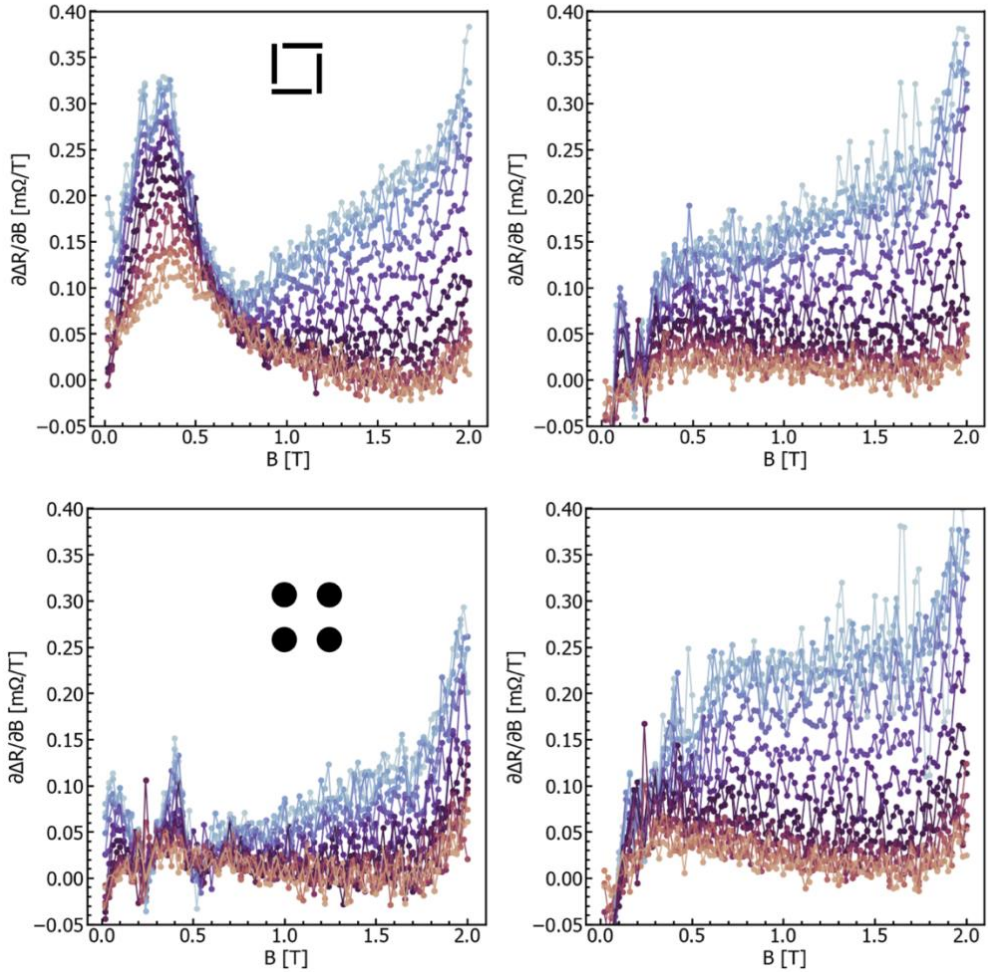

*Figure S4* **Asymmetric longitudinal transport response from chiral superlattice**. Comparison of the asymmetry in magnetic field of the resistance for the chiral (top) and square (bottom) superlattice devices. Left column is asymmetric response from the patterned side, whereas right is from the unpatterned side in each device. In the low field regime, only the chiral pattern shows a strong peak in the low field regime, where commensuration effects are present. This peak comes from CO effects being different for positive and negative magnetic fields.

**S5: Asymmetric response in the chiral superlattice from equations of motion.**

To derive the equations of motion for a charged particle subject to both a magnetic field in the *z*-direction and periodic potential, we begin with the Hamiltonian:

$$H = \frac{(\mathbf{p}-q\mathbf{A})^2}{2m} + U(x,y)$$

Where $q$ and $m$ are the particle charge and mass, respectively, $\mathbf{p}$ is the momentum, $\mathbf{A}$ is the vector potential, and $U(x,y)$ is the 2D superlattice potential. We use the symmetric gauge choice $\mathbf{A} = -\mathbf{r} \times \frac{\mathbf{B}}{2}$ following the derivation in Ref. [1] which results in the following equations of motion

$$\dot{x} = \frac{\partial H}{\partial p_x} = v_x$$

$$\dot{v}_x = -\frac{\partial H}{\partial x} = 2\sqrt{2}\, q \left(\frac{B}{B_0}\right) v_y - \frac{\partial U}{\partial x}$$

$$\dot{y} = \frac{\partial H}{\partial p_y} = v_y$$

$$\dot{v}_y = -\frac{\partial H}{\partial y} = -2\sqrt{2}\, q \left(\frac{B}{B_0}\right) v_x - \frac{\partial U}{\partial y}$$

Where $B_0 = 2(2mE_f)^{.5}$. In a square superlattice potential, $U(x,y)$ is symmetric under four-fold rotational symmetry, and mirror symmetry about both the *x* and *y* directions. Thus, the equations of motion are symmetric under the combined charge conjugation $C$ and parity $P$ operations. This can be seen by making the substitution $x \rightarrow -x$ followed by $q \rightarrow -q$.

Parity in the *x*- direction:

$$-\dot{x} = -v_x \rightarrow \dot{x} = v_x$$

$$-\dot{v}_x = 2\sqrt{2}\, q \left(\frac{B}{B_0}\right) + \frac{\partial U}{\partial x} \rightarrow \dot{v}_x = -2\sqrt{2}\, q \left(\frac{B}{B_0}\right) v_y - \frac{\partial U(-x,y)}{\partial x}$$

$$\dot{y} = v_y$$

$$\dot{v}_y = 2\sqrt{2}\, q \left(\frac{B}{B_0}\right) v_x - \frac{\partial U(-x,y)}{\partial y}$$

Thus, under the additional charge conjugation transformation $q \rightarrow -q$ and the fact that $U(x,y) = U(-x,y)$ the equations of motion are *CP* symmetric. Note that as an alternative to changing the sign of the charge $q$, the same symmetry is attained by switching the sign of the magnetic field $B \rightarrow -B$ and hence the equations of motion under positive and magnetic field should be the same.

In the case of a chiral superlattice potential, *U(x,y)* can still have rotational symmetry (four-fold for the case of the superlattice presented in the main text), but it has broken mirror symmetry along the *x*- and *y*- directions leading to $U(x,y) \neq U(-x,y)$ and $U(x,y) \neq U(x,-y)$. As a result, the equations of motion are manifestly *not* symmetric under the combined *CP* operation or the reversal of magnetic field direction.